\documentclass[aip,cha,reprint]{revtex4-1}

\usepackage{amsmath}
\usepackage{amssymb}  
\usepackage{algorithm2e}      
\usepackage{environ}  
\usepackage{graphicx}
\usepackage{hyperref} 
\usepackage[usenames,dvipsnames]{xcolor}

\newcommand{\eg}{\textit{e.g.}}

\newcommand{\ie}{\textit{i.e.}}

\newcommand{\eq}[1]{(\ref{eq:#1})}
\newcommand{\eqlabel}[1]{\label{eq:#1}}
\def\eqreftwo(#1,#2){(\ref{eq:#1},\ref{eq:#2})}
\newcommand{\eqtwo}[1]{\eqreftwo(#1)}
\def\eqrefthree(#1,#2,#3){(\ref{eq:#1},\ref{eq:#2},\ref{eq:#3})}
\newcommand{\eqthree}[1]{\eqrefthree(#1)}
\def\eqreffour(#1,#2,#3,#4){(\ref{eq:#1},\ref{eq:#2},\ref{eq:#3},\ref{eq:#4})}

\def\eqreffive(#1,#2,#3,#4,#5){(\ref{eq:#1},\ref{eq:#2},\ref{eq:#3},\ref{eq:#4},\ref{eq:#5})}

\NewEnviron{eqsplit}{\begin{equation}\begin{split} \BODY \end{split}\end{equation}}

\newcommand{\Fig}[1]{Fig.~\ref{fig:#1}}
\newcommand{\fig}[1]{fig.~\ref{fig:#1}}
\newcommand{\figlabel}[1]{\label{fig:#1}}

\newcommand{\sglfigure}[3]{\begin{figure}[tbp]\centerline{#1}\caption[]{#2}\figlabel{#3}\end{figure}}

\newcommand{\refcite}[1]{Ref.~[\onlinecite{#1}]}
\newcommand{\refcites}[1]{Refs.~[\onlinecite{#1}]}

\newcommand{\seclabel}[1]{\label{sec:#1}}
\newcommand{\secref}[1]{\ref{sec:#1}}
\newcommand{\Secn}[1]{Section~\secref{#1}}

\newcommand{\secn}[1]{Section~\secref{#1}}

\renewcommand{\@}{\partial}

\newcommand{\bydef}{\;\triangleq\;}

\newcommand{\conj}[1]{\overline{#1}}
\renewcommand{\d}{\mathrm{d}}

\newcommand{\Df}[2]{\dfrac{\d #1}{\d #2}}
\newcommand{\Ddf}[2]{\dfrac{\d^2{#1}}{\d{#2}^2}}
\newcommand{\df}[2]{\dfrac{\partial #1}{\partial #2}}
\newcommand{\ddf}[2]{\dfrac{\partial^2 #1}{\partial #2^2}}

\newcommand{\e}{\mathrm{e}}

\newcommand{\Heav}{\mathrm{H}}   

\newcommand{\inner}[2]{\left\langle #1 \,\Big|\, #2 \right\rangle} 
\newcommand{\intinf}{\int\limits_{-\infty}^{\infty}}

\newcommand{\Lerch}{\Omega}      

\newcommand{\Mx}[1]{\begin{pmatrix}#1\end{pmatrix}}
\newcommand{\mx}[1]{\mathbf{#1}}

\renewcommand{\O}[1]{\mathcal{O}\!\left(#1\right)} 

\newcommand{\Real}{\mathbb{R}}

\newcommand{\sech}{\mathrm{sech}}
\newcommand{\T}{^T}

\ifx01
\ifx\notcolor\undefined\newcommand{\notcolor}{blue}\else\fi
\else
\ifx\notcolor\undefined\newcommand{\notcolor}{black}\else\fi
\fi
\newcommand{\+}[2]{\def#1{{\color{\notcolor}#2}}}
\newcommand{\1}[2]{\def#1##1{{\color{\notcolor}#2}}}

\+{\0}{\mx{0}}      
\+{\A}{A}           
\+{\a}{a}           
\+{\ampu}{U\st}     
\+{\ampuc}{U_{\mathrm{s}}^*}  
\+{\ampi}{I\st}     
\+{\ampic}{I_{\mathrm{s}}^*}  
\+{\Cone}{C_1}      
\+{\Ctwo}{C_2}      
\+{\c}{c}           
\+{\co}{\c_0}       
\+{\ci}{\c_1}       
\+{\b}{b}           
\+{\bD}{\mx{D}}     
\+{\D}{D}           
\+{\Den}{\mathcal{D}}         
\+{\dim}{d}         
\+{\dt}{\Delta_t}   
\+{\dx}{\Delta_x}   
\+{\E}{E}           
\+{\Ec}{\hat{E}}    
\+{\Ep}{E_1}        
\+{\Er}{E_{\mathrm{r}}} 
\+{\best}{\mx{e}}   
\+{\est}{e}         
\+{\Est}{E}         
\+{\exty}{\tau}     
\+{\ff}{\mx{f}}     
\+{\f}{f}           
\+{\fal}{\alpha}    
\+{\fep}{\gamma}    
\+{\fth}{\beta}     
\+{\G}{g}           
\+{\lG}{G}          
\+{\g}{\mx{h}}      
\+{\h}{h}           
\+{\Ione}{I_1}      
\+{\Itwo}{I_2}      
\+{\Ithree}{I_3}    
\+{\Ifour}{I_4}     
\+{\Ifive}{I_5}     
\+{\i}{i}           
\+{\j}{j}           
\+{\k}{k}           
\+{\L}{\mathcal{L}} 
\+{\Lo}{\L_0}       
\+{\Li}{\L_1}       
\+{\Lp}{\L^{+}}     
\+{\Lpo}{\L^{+}_0}  
\+{\Lpi}{\L^{+}_1}  
\1{\LV}{\mx{W}_{#1}} 
\1{\LVo}{\widetilde{\mx{W}}_{#1}} 
\1{\LVi}{\widehat{\mx{W}}_{#1}}   
\1{\lw}{\rw{#1}}    
\+{\l}{\ell}        
\1{\lv}{W_{#1}}     
\1{\lvo}{\widetilde{W}_{#1}} 
\1{\lvi}{\widehat{W}_{#1}}   
\1{\lw}{\rw{#1}}    
\+{\N}{N}           
\+{\Num}{\mathcal{N}}         
\+{\Numone}{\mathcal{N}_1}    
\+{\Numtwo}{\mathcal{N}_2}    
\+{\n}{n}           
\+{\nn}{n}          
\+{\P}{P}           
\+{\p}{p}           
\+{\pf}{\nu}        
\1{\phir}{\phi_{#1}}   
\1{\phiro}{\widetilde{\phi}_{#1}} 
\1{\phiri}{\widehat{\phi}_{#1}}   
\1{\phil}{\phi^*_{#1}} 
\1{\philo}{\widetilde{\phi}^*_{#1}} 
\1{\phili}{\widehat{\phi}_{#1}^*}   
\1{\psir}{\psi_{#1}}   
\1{\psiro}{\widetilde{\psi}_{#1}} 
\1{\psiri}{\widehat{\psi}_{#1}}   
\1{\psil}{\psi^*_{#1}} 
\1{\psilo}{\widetilde{\psi}^*_{#1}} 
\1{\psili}{\widehat{\psi}^*_{#1}}   
\+{\q}{q}           
\+{\R}{R}           
\1{\RV}{\mx{V}_{#1}}     
\1{\RVo}{\widetilde{\mx{V}}_{#1}} 
\1{\RVi}{\widehat{\mx{V}}_{#1}}   
\+{\r}{r}           
\1{\rv}{V_{#1}}     
\1{\rvo}{\widetilde{V}_{#1}} 
\1{\rvi}{\widehat{V}_{#1}}   
\1{\rw}{\lambda_{#1}}    
\1{\rwo}{\widetilde{\lambda}_{#1}} 
\1{\rwi}{\widehat{\lambda}_{#1}}   
\+{\Speed}{S}       
\+{\Scale}{\mx{S}}  
\+{\shift}{s}       
\+{\sig}{\sigma}    
\+{\solvr}{f_e}     
\+{\solvl}{\tilde{f}_e} 
\+{\solvln}{\tilde{v}_e} 
\+{\st}{_{\mathrm{s}}} 
\+{\sRV}{\overline\RV}  
\+{\sLV}{\overline\LV}  
\+{\sprecomp}{\overline\precomp}  
\+{\Trone}{\Upsilon_1}  
\+{\Trtwo}{\Upsilon_2}  
\+{\rrr}{\rho}      
\+{\t}{t}           
\+{\tf}{\tau}       
\+{\tfs}{\tau'}     
\+{\tst}{\t\st}     
\+{\U}{U}           
\+{\u}{u}           
\+{\uc}{\hat{\u}}   
\+{\ucinner}{\uc_{inn}} 
\1{\uci}{\tilde\u_{#1}}   
\+{\ucouter}{\uc_{out}} 
\1{\uco}{\bar\u_{#1}}     
\+{\ucover}{\uc^{\natural}}
\+{\V}{V}           
\+{\v}{v}           
\+{\vc}{\hat{\v}}   
\+{\vcinner}{\vc_{inn}} 
\1{\vci}{\tilde\v_{#1}}   
\+{\vcouter}{\vc_{out}} 
\1{\vco}{\bar\v_{#1}}     
\+{\vcover}{\vc^{\natural}}
\+{\wc}{\hat{\w}}   

\+{\val}{\zeta}     
\+{\Xp}{\mx{X}}     
\+{\Xir}{\mx{\Xi}}  
\+{\Xil}{\tilde{\mx{\Xi}}} 
\+{\x}{x}           
\+{\xa}{x_*}        
\+{\xf}{\xi}        
\+{\xs}{\xi'}       
\+{\xt}{\xi''}      
\+{\xst}{\x\st}     
\+{\z}{z}           

\+{\w}{w}           

\+{\myxi}{\xi}     
\+{\bu}{\mx{u}}     
\+{\buc}{\hat{\bu}} 
\+{\bC}{\mx{C}}     

\+{\IRH}{I_{rh}}    
\+{\CHRO}{\tau}     

\+{\seq}{{\mu}}       

\+{\myzeta}{\zeta}

\begin{document}
\title{Fast-slow asymptotic for semi-analytical ignition
criteria in FitzHugh-Nagumo system}
\author{B. Bezekci}
\affiliation{College of Engineering, Mathematics and Physical Sciences, University of Exeter, Exeter EX4 4QF, UK}
\author{V. N. Biktashev}
\email[Corresponding author:]{V.N.Biktashev@exeter.ac.uk}
\affiliation{College of Engineering, Mathematics and Physical Sciences, University of Exeter, Exeter EX4 4QF, UK}

\begin{abstract}
  We study the problem of initiation 
  of excitation waves in the FitzHugh-Nagumo model. 
  Our approach follows earlier works and is based on the idea of 
  approximating the boundary between basins of attraction 
  of propagating waves
  and of the resting state as the stable manifold of a
  critical solution. 
  Here, we obtain analytical expressions for the essential ingredients
  of the theory by singular perturbation using two small parameters, the
  separation of time scales of the activator and inhibitor, and the
  threshold in the activator's kinetics. This results in a closed
  analytical expression for the strength-duration curve. 
\end{abstract}


\maketitle
\begin{quotation}
  Excitable reaction-diffusion systems underlie a large number of
  nontrivial spatio-temporal dynamic regimes and arise as models of a
  wide variety of physical, chemical and biological systems, some of
  which of considerable practical importance. One of such areas
  is electrophysiology of propagation of electric pulses in
  nerves and in the cardiac muscle. The detailed mathematical study of
  such models starts from \citet{hodgkin1952quantitative}. In their
  Nobel Prize work, they described how action potentials in neurons are
  initiated and propagated.  Due to the complexity of a four-variable
  system, a particular attention has been devoted to obtain simpler and
  more mathematically tractable systems, one of which is
  FitzHugh-Nagumo
  model~\cite{fitzhugh1961impulses,fitzhugh1960thresholds,nagumo1962active},
  widely accepted as an archetypical excitable model.  Conditions of
  existence of propagating waves in such models and their properties is
  a subject of vast research literature. However, the question of conditions
  required to initiate such waves, or factors that may quench them, are
  no less important for
  applications\cite{britton1982threshold,zipes2009cardiac}, and yet they are
  studied much less, because they are more complicated
  mathematically. The present paper is a part of an attempt to cover
  this gap, and endeavours to propose an analytical, albeit approximate,
   description of the initiation conditions, where previously only
   numerical treatment was believed possible. 
\end{quotation}

\section{Introduction}
\seclabel{intro}

Excitation waves may be defined as propagating nonlinear dissipative
waves in a system which also possesses a spatially uniform resting
state, stable with respect to small perturbations. Hence a transition
from the resting state to a propagating wave requires a sufficiently 
large perturbation. The problem of what perturbations are sufficient
to ignite an excitation wave is nonlinear, nonstationary, and
generally lacks any helpful symmetries, thus generally is considered
suitable only for numerical treatment. However, this problem has so
many important applications that any analytical answers, even if only
approximate and qualitative, are on high demand. 
One such analytical approach was investigated in 
our previous work~\cite{bezekci2015semianalytical,bbvnbists}.
It is based on linearization of the dynamic equations around so-called
critical solutions. These are unstable propagating waves (in some
cases, stationary ``nuclei'') which have exactly one unstable
eigenvalue, so their centre-stable manifolds serve as boundaries
separating the basins of attraction of the two possible outcomes,
ignition (generation of the propagating wave) and failure (return to
the resting state).
This approach has demonstrated viability on some examples, 
but has a disadvantage in 
that the essential
ingredients of the ignition criterion, such as the critical solution
itself, as well as its leading eigenvalues and eigenfunctions, are to
be obtained numerically. 
In this article, we focus particularly on the analytical
initiation criterion in a spatially extended FitzHugh-Nagumo system,
in which approximate propagating wave solutions are known in the limit
of two small parameters, the ratio of the characteristic times of the
activator and the inhibitor, and the threshold in the nonlinear
kinetics of the activator. We use singular perturbation theory to
construct the required ingredients for the linearized ignition
criterion, and see how well the resulting criterion works. 

The structure of the paper is as follows. In \Secn{anatheo}, the
analytical theory proposed in the earlier publications is summarized,
with application to a
two-component test problem, the FitzHugh-Nagumo model. The main
contribution of this study is the analytical derivations of the
essential ingredients of the threshold curves by means of the
perturbation theory.  \Secn{pertana} shows how these ingredients are
obtained, along with the strength-duration curve
approximation. Finally, in \Secn{disc} a short discussion of the results
and some possible further research will be given.

\section{Theory}
\seclabel{anatheo}

The FitzHugh-Nagumo model (FHN) may be presented in various
  equivalent forms. For this paper, we prefer the formulation used
  \eg\ by \textcite{Neu1997}:
\begin{align}
  & 
  \u_\t=\u_{\x\x}+\f(\u)-\v, \qquad
  \nonumber\\&
  \v_\t=\fep(\fal\u-\v), 
  \eqlabel{FHN}
\end{align}
where $\f(\u)$ is a cubic polynomial function in the form $\f(\u)=\u\left(\u-\fth\right)\left(1-\u\right)$, the variables $\u$ and  $\v$ represent respectively membrane potential and the  recovery variables,
$\fep$  is a small parameter describing the ratio of time scales of the variables $\u$ and $\v$, and $\fal$ is a constant.
The parameter $\fth$ plays a key role in the fast dynamics of the model as it is the threshold state of the system that must be in the range $\left(0,1/2\right)$  in order for the system to have a qualitative electro-physiological meaning~\cite{maginu1978stability,maginu1980existence}. 

In this paper, we consider the problem of initiation by
  a current pulse, modelled as a non-homogeneous Neumann boundary
  condition, with the rectangular profile of duration 
  $\tst$ and strength of the current $\ampi$,
\begin{align}
  \u_\x(0,\t)=-\ampi\Heav(\tst-\t), \quad \v_\x(0,\t)=0,\quad \t>0,  \eqlabel{semi-cable}
\end{align}
where $\Heav(\cdot)$ is the Heaviside step function. 
%
%
Asymptotic
behaviour of the solutions of excitable reaction-diffusion
  systems has been a topic of
intense study, see for
example~\refcites{aronson1975nonlinear,mckean1985threshold,flores1989stable}. Typically,
the solution of \eq{FHN} either approaches the propagating pulse
solution (``ignition'') or the resting state (``failure''). 
A curve in the $(\tst,\ampi)$-plane
that separates initial conditions leading to
the ignition and initial conditions leading to the resting state is
called a strength-duration curve. We shall also refer to
it as a ``threshold curve'', or ``critical curve''. 

\sglfigure{\includegraphics{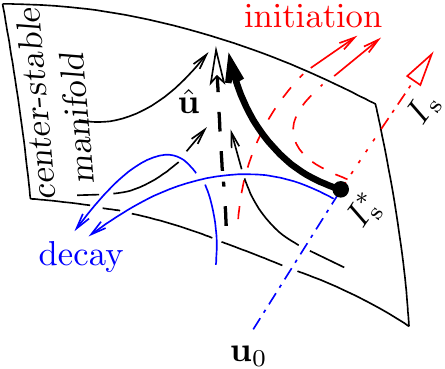}}{ %
  The sketch of a center-stable manifold of a moving critical
  solution. The dashed black line denotes the critical solution
  $\bu=\buc(\x-\c\t)$, while
  the solid black lines represent the critical trajectories that form
  the center-stable manifold. The bold solid black line is the critical
  trajectory that divides the family of initial conditions into two
  classes: sub-threshold trajectories (blue lines) and super-threshold
  trajectories (red lines). The dash-dotted line represents the
  one-parametric family of initial
  conditions $\bu_0:=\bu(\x,\tst)$, parameterized  by $\ampi$.
  The point where the curve of initial
  conditions intersects the center-stable manifold at the threshold
  value $\ampic$ of the parameter is shown as the
  filled circle. %
}{censtab}

The mathematical description of the threshold curve is motivated by the
existence of a ``critical solution'', which is an unstable
  relative equilibrium with a single unstable eigenvalue; for the
  FitzHugh-Nagumo system~\eq{FHN} it is a ``critical
  pulse''~\cite{flores1989stable,flores1991stabilitynew,idris2008analytical,idris2007critical,burhan2016thesis}. 
The
stable manifold of such critical solution has codimension two, whereas
its center-stable manifold has codimension one and as such, it can
partition the phase space into two basins of attraction: one corresponds
to the decay solutions, and the other to the initiation solutions, as
sketched in \Fig{censtab}. Of course, the concept of the
  ``basin of attraction'' is applicable to an autonomous problem,
  whereas our problem \eqtwo{FHN,semi-cable} is not. However, given a
  finite duration of the stimulating current, $\tst$, we have an
  autonomous system for all $\t\ge\tst$, so the outcome of a
  stimulation will depend on which side of the centre-stable manifold will
  $\bu_0(\x)=\bu(\x,\tst)$ be. This is further simplified by the fact
  that, even for a fixed $\tst$, we have a one-parameter family of
  such initial conditions, depending on the parameter $\ampi$. This
  will correspond to a curve $\bu_0(\x;\ampi)$ in the functional
  space (dash-dotted line in~\fig{censtab}). The critical value
  $\ampic$, corresponding to the boundary between failure and success,
  corresponds to the point of  intersection  between this curve and the
  centre-stable manifold (bold filled circle). By construction, the
  initial protocol corresponding to the intersection point, produces
  a trajectory (bold arrow) that lies on the boundary of the basins, \ie\ is a
  ``saddle straddle trajectory'' in the terminology of~\refcite{Nusse-Yorke-1989}, and approaches
  the unstable solution $\buc$  (dashed arrow) as $\t\to\infty$.

Analytical expressions for the ignition criteria can be obtained by
approximating this center-stable manifold by its tangent at the critical
solution, \ie\ the center-stable space; the feasibility of quadratic
approximation was also demonstrated in some cases. Details of this
approach have been described
elsewhere\cite{bezekci2015semianalytical,bbvnbists}, and here we only
quote the required results. The primary ingredient for the theory is of
course the critical solution itself, which for the FHN system has the
form of the critical pulse
$\bu(\x,\t)=\buc(\xf)=\Mx{\uc(\xf), \vc(\xf)}\T$, %
where
$\xf=\x-\c\t-\shift$, %
the constant (nonlinear eigenvalue)
$\c$ is the speed of the critical pulse, and $\shift\in\Real$ is an
arbitrary constant. %
Beyond that, the theory requires two leading left eigenfunctions %
$\LV1=\Mx{\phil1,\psil1}\T$, $\LV2=\Mx{\phil2,\psil2}\T$ %
and first leading eigenvalue $\rw1$; note that $\rw2=0$, due to the
translational symmetry. The ignition criterion has been formulated as a
finite nonlinear system of two equations for two unknowns, $\shift$ and
$\ampi$,
\begin{align}
  \begin{cases}
2\ampi\int\limits_0^{\tst}\e ^{-\rw1 \tfs}\phil1\left(-\c\tfs-\shift\right)\, \d {\tfs} & = \Numone, \\
2\ampi\int\limits_0^{\tst}\phil2\left(-\c\tfs-\shift\right)\,  \d {\tfs}  & = \Numtwo,
  \end{cases}                                     \eqlabel{crit-sys}
\end{align}
where the right-hand sides $\Numone$ and $\Numtwo$ are constants, defined entirely
by the properties of the model,
\begin{align}
  \Num_\l=\inner{\Mx{\phil\l(\xf)\\\psil\l(\xf)}}{\Mx{\uc(\xf)\\\vc(\xf)  }},  \qquad  \l=1,2.        \eqlabel{Numnboth}
\end{align}
Here and below, we use the bra-ket notation for the inner product: if
$\mx{v}=\Mx{a,b}\T$ and $\mx{w}=\Mx{c,d}$, then
\[
  \inner{\mx{w}}{\mx{v}} \bydef \intinf \left(
    \conj{c}a + \conj{d}b
  \right) \,\d\xf. 
\]

The compatibility condition for the two equations given in \eq{crit-sys} for $\ampi$ is 
\begin{align}\nonumber
& \seq(\shift) \bydef
\Numone\int\limits_{-\shift}^{-\c \tst-\shift} 
\phil2\left(\val\right)\, \d  \val 
 \nonumber \\
& 
-\Numtwo\e^{\rw1 \shift/\c}\int\limits_{-\shift}^{-\c \tst-\shift} \e^{\rw1 \val/\c}
\phil1\left(\val\right)\,\d \val=0.
\eqlabel{zeroofs}
\end{align}
In the previous works, we have been unable to find any ingredients in
these formulations required for the definition of strength-duration
threshold curve analytically with a few exceptions of limited practical
importance. Hence, a hybrid approach where these key ingredients are
determined numerically must typically be employed. In the present paper,
we obtain the required ingredients analytically, which will allow
description of the strength-duration curve in a closed analytical
form. This is achieved by using perturbation theory with $\fep$ and
$\fth$ as small parameters.

\section{Perturbation Analysis}
\seclabel{pertana}

In this section, we employ the perturbation theory  to
obtain the critical pulse and leading eigenfunctions and corresponding
eigenvalues of FHN system using the exact solution of its fast subsystem,
Zeldovich-Frank-Kamenetsky (ZFK) equation, sometimes also called Nagumo
equation:
\begin{align} \eqlabel{ZFK}
  \u_\t=\u_{\x\x}+\f(\u) . 
\end{align}
Clearly, when we set $\fep=0$ and $\v\equiv 0$, FHN system transforms
into ZFK equation and formally, we use a series in $\fep$ and the solution of
ZFK equation to have the approximation to the full solution of FHN system.

\subsection{Finding the critical pulse}

To find the critical pulse, we look for solutions of 
the form $\u(\x,\t)=\uc(\xf)$, $\v(\x,\t)=\vc(\xf)$ where $\xf=\x-\c\t$, and 
the positive constant $\c$ is the propagation speed of the rightward traveling wave, 
yet to be determined. Then, the FHN system is converted into following system of 
first order ordinary differential equations:
\begin{align}
& \uc_{\xf} = \wc, \nonumber \\
& \wc_{\xf} = \c\wc-\f(\uc)+\vc,  \eqlabel{eqpertch6} \\
& \vc_{\xf} = \frac{ \fep(\fal\uc-\vc)}{\c}. \nonumber
\end{align}
The traveling wave solution vanishes at both infinities along with its
first derivative
\begin{align}  \eqlabel{pulseBC}
 \lim_{ |\xf|\to\infty}  \uc= \lim_{ |\xf|\to\infty}
   \vc =\lim_{ |\xf|\to\infty}  \wc=0.   
\end{align}
\citet{hastings1976existence} has proved that for a sufficiently small
$\fep$, there are at least two distinct positive numbers $\c$ such
that the above system has a homoclinic.  It was also proved that the
higher speed is
$\c\left(\fep\right)\approx\sqrt{2}\left(\frac{1}{2}-\fep\right)$ and
the corresponding pulse is stable, while the slower speed is
$\c\left(\fep\right) = \O{\sqrt{\fep}}$ and the slower pulse is
unstable, with one positive
eigenvalue~\cite{flores1991stabilitynew,hastings1982single}.  That is,
the slow pulse is our critical pulse, and we restrict our analysis to
it. We use the matching asymptotics method: divide the domain of the
problem into two subdomains: the inner region where the solution
changes rapidly ((with a speed $\O{1}$) in the limit $\fep\searrow0$,
and the outer solution where it varies slowly (with a speed
$\O{\fep^{1/2}}$). The solutions obtained in these regions are called
as inner and outer solutions, respectively. The inner and outer
solutions are then matched to ensure that the approximate solution is
uniformly valid in the whole domain. This is achieved by using a
transition zone, in which the two solutions are asymptotically equal.

\subsubsection{Inner Expansion}

Following ~\refcite{casten1975perturbation}, we represent the inner
asymptotics of the solution in the form
\begin{align}
&  \ucinner(\xf) = \uci0(\xf)+ \fep^{1/2} \uci1(\xf)+\fep \uci2(\xf)+\ldots  , \nonumber \\
&  \vcinner(\xf) = \fep^{1/2} \vci1(\xf)+\fep \vci2(\xf)+\ldots  ,  \eqlabel{parpert} \\
&  \c = \fep^{1/2} \ci+\ldots, \nonumber 
\end{align}
where $\vci0\equiv 0$ and $\co=0$ as the ZFK equation is one-component and its critical 
nucleus solution has zero velocity. Substituting these into equation \eq{eqpertch6} and 
collecting the terms by the powers of $\fep^{1/2}$, we have 
\begin{align}
& \Ddf{\uci0}{\xf}+\f\left(\uci0\right)=0,  \eqlabel{pert1ch6}  \\
& \Ddf{\uci1}{\xf}+\f'\left(\uci0\right)\uci1-\ci\Df{\uci0}{\xf}-\vci1=0, 
 \eqlabel{pert2u1eqch6}  \\
& \Df{\vci1}{\xf}- \frac{\fal\uci0}{\ci}=0.  \eqlabel{pert3veqch6} 
\end{align}
Coefficient $\ci$ can be determined by multiplying \eq{pert2u1eqch6} 
by $\d \uci0/\d\xf$  and then integrating the result using the 
equations \eq{pert1ch6} and \eq{pert3veqch6} therein, giving
\begin{align}
\ci=\left(\frac{\fal \intinf \uci0^2\, \d \xf}{\intinf \left(\frac{\d \uci0}{\d \xf}\right)^2\,\d \xf} \right)^{1/2}, 
\end{align}
where the integration is performed over the inner domain. As we aim to
obtain explicit analytical expressions as much as possible, we
consider the limit of small $\fth$.  For $\u\lesssim \fth$ we can
approximate $\f(\u)\approx \u\left(\u-\fth\right)$ for which the
critical nucleus of ZFK is\cite{idris2008analytical,Neu1997}
\begin{align*}
	\uci0=3 \fth\, \sech^2\left(\xf\sqrt {\fth}/2\right)/2, 
\end{align*}
so that the speed correction evaluates to 
\begin{align*}
  \ci=\sqrt {{\frac {5\fal}{\fth}}}. 
\end{align*}
Equation \eq{pert3veqch6} for $\vci1$ is first order separable,
and its solution satisfying $\vci1(-\infty)=0$ is
\begin{align}
 \vci1  = 3\fth\sqrt{\fal/5} \left[
   1+\tanh\left(\xf\sqrt{\fth}/2\right)
  \right] .
\end{align}
Equation \eq{pert2u1eqch6} for $\uci1$ is second order linear and we
know that $\d\uci0/\d\xf$ is a solution, so its general solution can
be found by the substitution
\begin{align*}
	\uci1=\p(\xf)\frac{\d\uci0 }{\d\xf},
\end{align*}
which gives a first-order linear equation for $\p$, leading to
\begin{align}
 \uci1= & 
    \frac{
      -6\sqrt\fal\,\e^{\xf\sqrt\fth}
    }{
      \sqrt{5\fth}\left(\e^{\xf\sqrt\fth} + 1\right)^3
    } \\ \nonumber
  & \times \left[
    \e^{2\xf\sqrt\fth} \sqrt\fth
    +\e^{\xf\sqrt\fth} \left(3\fth\xf - 4\sqrt\fth\right)
    -3\fth\xf - 5\sqrt\fth 
  \right].
\end{align}

\subsubsection{Outer Expansion}

Technically, we have two outer regions, for $\xf<0$ and $\xf>0$. The
inner solution obtained above satisfies the boundary conditions at
$\xf\to-\infty$, but not at $\xf\to\infty$. Hence the outer solution
for $\xf<0$ may be taken as zero both for $\u$ and $\v$, whereas for
$\xf>0$ a nontrivial solution must be found.  There, we use the
independent variable $\myzeta=\fep^{1/2}\xf$, and assume the solution
in the form of a power series in $\fep^{1/2}$, starting with
\begin{align}
& \ucouter= \fep^{1/2} \uco1\left(\myzeta\right)+\ldots, \nonumber \\
& \vcouter= \fep^{1/2} \vco1\left(\myzeta\right)+\ldots .
\end{align}
Substituting these into \eq{eqpertch6} and collecting the terms by the
powers of $\fep^{1/2}$, we get
\begin{align}
& \fth\uco1+\vco1 =0, \\
& \ci\frac{\d \vco1}{\d \myzeta}=\fal \uco1-\vco1,
\end{align}
which have the following nontrivial solutions 
\begin{align}
& \vco1( \myzeta)=  \A\,{{\e}^{{-\myzeta \left( {\frac {\fal+\fth}{
\sqrt{5\fal\fth}}} \right)}}},\\ 
& \uco1( \myzeta)= -\frac {\A}{\fth} \,{{\e}^{{-\myzeta \left( {\frac {\fal+\fth}{
\sqrt{5\fal\fth}}} \right)}}},
\end{align}
where the constant $\A$ is to be determined from the condition that the
inner and outer expansions give the same result in the transition
zone. This can be achieved by applying the Van Dyke's matching
principle~\cite{van1975perturbation}, requiring that the inner
solution in the transition zone (\ie~as $\xf \to \infty$) is equal to
the outer solution in the transition zone (\ie~as $\myzeta\to 0$):
\begin{align}\eqlabel{vanconch6}
\lim_{\xf\to\infty} \vci1 = \lim_{\myzeta\to 0} \vco1,
\end{align}
which gives
\begin{align*}
   \A= \frac{6\fth\sqrt{\fal}}{\sqrt{5}}. 
\end{align*}

\sglfigure{\includegraphics{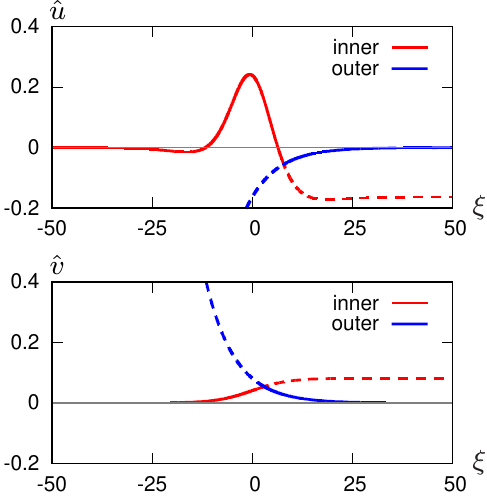}}{%
  Inner and outer solutions of $\uc$ (top) and $\vc$ (bottom)
  components of the perturbed critical pulse. Parameters used:
  $\fep=0.01$, $\fth=0.05$, $\fal=0.37$. %
}{fhn-innerouter}

\Fig{fhn-innerouter} shows the inner and outer solutions of $\uc$ and
$\vc$ components of the critical pulse for the FHN system for the
parameter values, $\fep=0.01$, $\fth=0.05$, $\fal=0.37$.  In the
negative $\xf$ region, the inner solution is uniformly valid. In the
positive $\xf$ region, on the other hand, neither the inner nor the
outer solutions alone can be the solution and we would like to combine
these two solutions into a ``composite solution'' that would be
uniformly valid. This can be done by adding the inner and outer
approximations and subtracting the matching value, which would have
been taken into account twice otherwise. Thus, our final critical
pulse solution based on perturbation theory, valid in the whole
domain, is in the following form,
\begin{align*}
& \uc(\xf)= \ucinner(\xf) + \left[
    \ucouter(\xf\sqrt\fep)-\ucover
  \right]\Heav(\xf), \\
& \vc(\xf)=
    \vcinner(\xf) + \left[
      \vcouter(\xf\sqrt\fep)-\vcover
  \right]\Heav(\xf),
\end{align*}
where the matching values are
\begin{align*}
  \ucover=& \ucinner(\infty) = \ucouter(0)  
            = - 6\sqrt{\fal\fep/5},
  \\
  \vcover=& \vcinner(\infty) = \vcouter(0) 
            = 6\fth\sqrt{\fal\fep/5},
\end{align*}
and we have dropped the terms $\O{\fep}$ throughout.

\sglfigure{\includegraphics{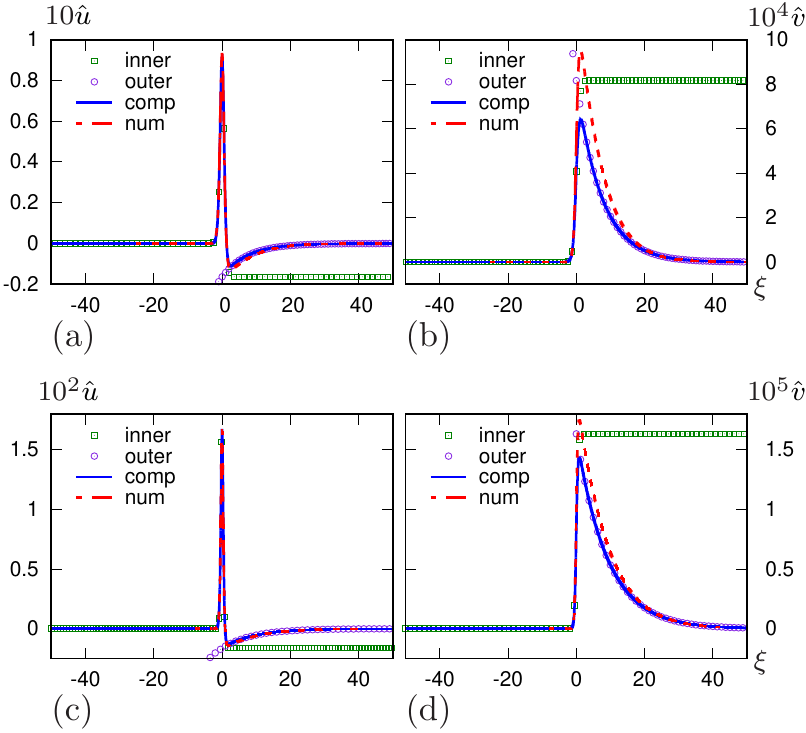}}{%
  Asymptotic critical pulses (inner solution, outer
    solution, `comp' for composite solution) compared to the ones
    obtained numerically (`num').
   Two sets of parameter values are chosen:
  $\fep=0.001$, $\fth=0.05$, $\fal=0.37$ (a,b) and $\fep=0.00001$,
  $\fth=0.01$, $\fal=0.37$(c,d). %
}{critpert}
 
\Fig{critpert} shows the critical pulse solutions of FHN system based
on the asymptotic perturbation theory analysis compared with the ones
obtained numerically. We used the numerical methods described
in~\refcite{bezekci2015semianalytical}. The figure illustrates two
selected set of parameters, $\fep=0.001$, $\fth=0.05$, $\fal=0.37$
(top panel) and $\fep=0.00001$, $\fth=0.01$, $\fal=0.37$ (bottom
panel). It can be seen that the asymptotic result gets closer to the
numerical critical pulse when the parameters $\fth$ and $\fep$ both
become smaller, which is indeed expected.

\subsection{Finding the Leading Eigenvalues  and Eigenfunctions}

In a similar fashion, perturbation theory can be applied to
approximate the eigenfunctions and eigenvalues of FHN
system. Generally speaking, the eigenvalue problem is to be solved by
matching asymptotics as well. However, in our case the inner solution
happens to vanish at both infinities so both outer solutions can be
taken as zero, and we only need to look at the inner solution.

To begin with, we linearize the FHN system \eq{FHN} around the
critical pulse in the co-moving frame of reference $\xf=\x-\c\t$,
$\tf=\t$,
\begin{align*}
 \u(\xf,\tf) =\uc(\xf)+\U\left(\xf,\tf\right), \quad   \v(\xf,\tf) =\vc(\xf)+\V\left(\xf,\tf\right),
\end{align*}
such that FHN system with quadratic nonlinearity gives
\begin{align}
  & \df{\U}{\tf} =\ddf {\U}  \xf- \c \df {\U} \xf +  \left(2\uc-\fth\right) \U- \V , \\  
  & \df{\V}{\tf} = -\c \df { \V}  \xf + \fep \left( \fal  \U  - \V\right).
\end{align}
We are looking for solutions of the linearized problem of the form
$\U\left(\xf,\t\right)= \e^{\rw{}\tf}\phir{}(\xf)$ 
and $ \V\left(\xf,\tf\right)= \e^{\rw{}\tf}\psir{}(\xf),$
which leads to the right eigenfunction problem, 
\begin{align} \eqlabel{orieigprobch6}
\rw{}\RV{}=\L \RV{},
\end{align}
where
\begin{align*}
\L = \Mx{ \partial^2_{\xf}-\c \partial_\xf+2\uc-\fth  & -1 \\
 \fep\fal  & -\c \partial_\xf-\fep }, \quad  \RV{}=\Mx{ \phir{} \\ \psir{}  }.
\end{align*}
Inserting the speed and critical pulse defined in the inner expansion analysis into 
the operator  $\L$, we have
\begin{align*}
  \L=\Lo+\fep^{1/2}\Li+\O{\fep},
\end{align*}
where
\begin{align*}
  & \Lo = \Mx{ \partial^2_{\myxi}+2\uci0-\fth  & -1 \\ 0  & 0 }, \\
  & \Li = \Mx{ -\ci\partial_\myxi+2\uci1  & 0 \\ 0  & -\ci\partial_\myxi }.
\end{align*}
Now, we expand the eigenvalues and eigenfunctions in a power series in terms 
of $\fep^{1/2}$,
\begin{align*}
  & \rw{} = \rwo{} + \fep^{1/2}\rwi{} + \O{\fep}, \\
  & \RV{} = \RVo{} + \fep^{1/2}\RVi{} + \O{\fep}.
 \end{align*} 
This kind of expansion has been widely used in the field of quantum mechanics, see for 
example \refcite{sakurai2011modern}. Implementing this eigenpair expansion into the original 
eigenvalue problem \eq{orieigprobch6}, we have
\begin{align*}
  & \left(\rwo{}+\fep^{1/2}\rwi{}\right)\left(\RVo{}
 +\fep^{1/2}\RVi{}\right) \\
 & = \left(\Lo+\fep^{1/2}\Li\right)
 \left(\RVo{}
 +\fep^{1/2}\RVi{}\right) +\O{\fep}. 
\end{align*}
Equating this in terms of the coefficients of the powers of $ \fep^{1/2}$, we get
\begin{align}
 & \O{\fep^0}: && \rwo\j \RVo\j=\Lo \RVo\j, 
 \eqlabel{idrispapL} \\
 & \O{\fep^{1/2}}: && \rwo\j \RVi\j+\rwi\j \RVo\j=\Li\RVo\j+\Lo\RVi\j, \eqlabel{perteqpart1L} 
\end{align}
for $\j=1,2,\ldots$. 
The leading order equation \eq{idrispapL} reduces to
the eigenvalue problem of the unperturbed problem, ZFK equation, and the leading eigenpair
for it is known explicitly~\cite{idris2008analytical}:
\begin{align*}
\RVo1
  =\Mx{ \phiro1 \\ \psiro1 }
  =\Mx{ \sech^3\left(\xf \sqrt{\fth}/2 \right)  \\ 0 },
  \quad \rwo1=\frac{5\fth}{4}. 
\end{align*}
Also, due to the translational symmetry, the second eigenpair is delivered by
the derivative of the critical nucleus solution of the ZFK equation,
\begin{align} \eqlabel{shiftmode0}
  \RVo2=\Mx{ \phiro2  \\ \psiro2 }=\Mx{ \d{\uci0}/\d{\xf}  \\ 0  }, \quad \rwo2=0.
\end{align}

As discussed in~\secn{anatheo}, the linearized ignition criterion
  requires knowledge of solutions of the
adjoint linearized problem
\begin{align} \eqlabel{adjlinprobfhnch6}
  \lw{} \LV{} = \Lp \LV{},  
\end{align}
where in our case
\begin{align*}
   \Lp = \Mx{ \partial^2_{\xf}+\c \partial_\myxi+2\uc-\fth  & \fep\fal \\ -1 & \c \partial_\xf-\fep }, \quad  \LV{}=\Mx{ \phil{}  \\ \psil{}  }. 
\end{align*}
We write
\begin{align*}
  \Lp=\Lpo+\fep^{1/2}\Lpi+\O{\fep},
\end{align*}
where
\begin{align*}
  \Lpo &= \Mx{ \partial^2_{\xf}+2\uci0-\fth  & 0 \\ -1  & 0 }, \\
  \Lpi &= \Mx{ \ci\partial_\xf+2\uci1  & 0 \\ 0  & \ci\partial_\xf }.
\end{align*}
We look for the left eigenfunctions in the
form of asymptotic series as
\begin{align*}
 \LV{}=\LVo{}+\fep^{1/2}\LVi{}+\O{\fep},
\end{align*}
and the series for the eigenvalues the same as for the right eigenfunctions. Inserting these 
series into \eq{adjlinprobfhnch6} and balancing the terms up to the order of $\fep^{1/2}$, we have 
\begin{align} 
 & \rwo\j \LVo\j=\Lpo \LVo\j, \eqlabel{idrispap} \\
 & \rwo\j \LVi\j + \rwi\j \LVo\j = \Lpi\LVo\j + \Lpo\LVi\j . \eqlabel{perteqpart1}
\end{align}
The leading order equation~\eq{idrispap} gives straightforwardly  
\begin{align*}
  & \LVo1=\Mx{\philo1 \\ \psilo1} = \Mx{ \phiro1 \\ -\phiro1/\rwo1}. \\
\end{align*}

Our next goal is to find the eigenvalue perturbations. We rewrite \eq{perteqpart1L} as
\begin{align} \eqlabel{lamperfound}
\left(\rwo\j-\Lo\right) \RVi\j
=
\left(\Li-\rwi\j\right) \RVo\j. 
\end{align}
We already know the leading order for $\j=1,2$. Now we take
the inner product of  the left-hand side of
\eq{lamperfound} with $\LVo\j$ to get
\begin{align*}
\inner{\LVo\j}{\left(\rwo\j-\Lo\right)\RVi\j} 
 =
    \rwo\j\inner{\LVo\j}{\RVi\j}
    -
    \inner{\Lpo \LVo\j} {\RVi\j} & \\
 =\rwo\j\inner{\LVo\j}{\RVi\j}
 -\rwo\j\inner{\LVo\j}{\RVi\j}=0, &
\end{align*}
since $\LVo\j$ is an eigenfunction of $\Lpo$ and the inner product is
semilinear in the first factor. Using this result in
\eq{perteqpart1L}, we obtain the classical expression for the
eigenvalue perturbations, 
\begin{align} \eqlabel{rew-1}
\rwi\j=\frac{ \inner{\LVo\j}{\Li\RVo\j}}
{\inner{\LVo\j}{\RVo\j}} .
\end{align}
In particular, for $\j=1$, we find the leading eigenvalue as
\begin{align} \eqlabel{rwi1}
 \rwi1 =\frac{\intinf \phiro1\left(-\ci\partial_\myxi+2\uci1\right)
 \phiro1\,\d \xf} {\intinf \phiro1^2\,\d \xf} = \frac{3\sqrt{5\fal}}{2},
\end{align}
and we of course have 
\begin{align} \eqlabel{rwi2}
  \rwi2=0
\end{align}
due to the translational symmetry. 

The linear approximations of the critical curves require the knowledge
of the left eigenfunctions, \ie~the eigenfunctions of the adjoint
linearized equation. Hence, we skip the details of the analytical
construction of the right eigenfunctions and proceed straight to finding the
left eigenfunctions perturbations, $\LVi\j$.  We begin with the first
component of the $\j=1$ left eigenfunction. It satisfies
\begin{align} \eqlabel{nonhomogeneousfhnch6}
  \left(\partial^2_{\xf}+2\uci0-\fth-\rwo1\right)\phili1=
  \left(\rwi1-c_1\partial_\xf-2\uci1\right)\philo1  
\end{align}
or
\[
  \phili1 '' +\P(\xf)\phili1=\R(\xf)
\]
where
\[
  \P(\xf)=2\uci0-\fth-\rwo1, 
  \quad
  \R(\xf)=\left(\rwi1-c_1\partial_\xf-2\uci1\right)\philo1.
\]
We know one solution of the corresponding homogeneous equation,
$\R(\xf)=0$, which is $\philo1= \sech^3\left(\xf \sqrt{\fth}/2 \right)$. So we
look for the solution of the full non-homogeneous equation using the
reduction-of-order substitution 
\begin{align*}
 \phili1(\xf)=\philo1(\xf)\pf(\xf), 
\end{align*}
leading to
\begin{align*}
  \pf''+\frac{2\philo1'}{\philo1}\pf'=\frac{\R}{\philo1} .
\end{align*}
The general solution of this is
\begin{align*}
 \pf(\xf)=\int\limits_{-\infty}^{\xf} \frac{\int\limits_{-\infty}^{\xs}
  \philo1(\xt) \R(\xt)\, \d \xt  +\Cone}{\left( \philo1 
  (\xs)\right)^2 }\,\d \xs  +\Ctwo, 
\end{align*}
where $\Cone$ and $\Ctwo$ are constants of the integration. 
The resulting explicit expression for $\phili1$ is
\begin{align} \eqlabel{phi1pertch6}
  \phili1(\xf)=&\philo1(\xf) \pf(\xf) =\sech^3\left(\xf\sqrt {\fth}/2\right)
  \left(\Ctwo+\Ione +\Cone \Itwo  \right)
\end{align}
where 
\begin{align*}
\Ione&=-{\frac {9\sqrt {\fal}\xf{{\e}^{\xf\sqrt {\fth}}}}{
\sqrt {5\fth} \left( {{\e}^{\xf\sqrt {\fth}}}+1 \right) }}, \\
\Itwo&=\frac{{\e}^{-3\,\xf\sqrt {\fth}}}{192\,\sqrt {
\fth}}\left( {
{\e}^{6\,\xf\sqrt {\fth}}}+9\,{{\e}^{5\,\xf\sqrt {\fth}}}+60\,\xf 
\sqrt {\fth}{{\e}^{3\,\xf\sqrt {\fth}}} \right. \\ & \left. +45\,{{\e}^{4\,\xf\sqrt {
\fth}}}-45\,{{\e}^{2\,\xf\sqrt {\fth}}}-9\,{{\e}^{\xf\sqrt {
\fth}}}-1
\right). 
\end{align*} 
Remember that the inner solution for the critical nucleus is uniformly
valid for $\xf<0$. Hence we expect the inner solutions for the
eigenfunctions also to be uniformly valid for $\xf<0$. In particular,
they should be bounded at $\myxi\to-\infty$. We have
\begin{align*}
 & \lim_{\xf\to-\infty} \sech^3\left(\xf\sqrt {\fth}/2\right)=0, \\
 & \lim_{\xf\to-\infty} \Ione\,\sech^3\left(\xf\sqrt {\fth}/2\right)=0, \\
 & \lim_{\xf\to-\infty} \Itwo\,\sech^3\left(\xf\sqrt {\fth}/2\right)=-\infty. \\
\end{align*}
Consequently, the constant $\Cone$ must be zero. Then \eq{phi1pertch6} simplifies to 
\begin{align}
\phili1(\xf)=\sech^3\left(\xf\sqrt {\fth}/2 \right) \left( \Ctwo+\Ione \right) .
\end{align}
Before we can find the constant $\Ctwo$, we need to find also the second 
component of the first adjoint eigenfunction; this is easily found as
\begin{align}
& \psilo1= -\frac{\philo1}{\rwo1}=
{\frac {-4 \sech^3\left(\xf\sqrt {\fth}/2\right)
 }{5\fth}} . \eqlabel{pertw1fhnch6}
\\
& \psili1=\frac{\ci \psilo1'  -\phili1 - \rwo1
\philo1}{\rwi1}. \eqlabel{pertw1fhnch6p2}
\end{align}
The value of the constant $\Ctwo$ can then be found from the orthogonality 
condition $\inner{\LV1}{\RV2}=0$. In the expanded form, this can be written as
\begin{align*} &
\inner {\Mx{\philo1  \\ \psilo1  }}{\Mx{ \uci0'  \\ 0  }}
\\ &
+\fep^{1/2}\left\{
\inner {\Mx{ \phili1  \\ \psili1  }}{\Mx{ \uci0'  \\ 0  }}
+
\inner {\Mx{ \philo1  \\ \psilo1  }}{\Mx{ \uci1'  \\ \vci1' }}
\right\} 
\\ &
+\fep
\inner {\Mx{\phili1  \\ \psili1  }}{\Mx{ \uci1' \\ \vci1' }}  =0,
\end{align*}
where the first term vanishes due to ~\eq{shiftmode0} as $\Mx{\uci0', & 0}\T$ is
the $\j=2$ right eigenfuntion so automatically is orthogonal to
$\Mx{\philo1,\psilo1}\T$.
Similarly, the second term 
in the above formula also vanishes since the sum of all three definite integrals is 
equal to zero as calculated below,
\begin{align*}
   \intinf \uci0' \phili1 \,\d\xf = 
  & \Ctwo\intinf \uci0' \philo1\, \d\myxi+
     \intinf \uci0' \philo1 \Ione\, \d\myxi
    \\
  = & \frac{81\pi\sqrt{5\fal}}{200}, 
\end{align*}
(the integral multiplied by $\Ctwo$ here vanishes as it is the same as
the one discussed above),
\begin{align*}
   & \intinf \uci1' \philo1\, \d \myxi= -\frac{9\pi\sqrt{5\fal}}{40}, 
\end{align*}
\begin{align*}
      & \intinf \vci1' \psili1\, \d \myxi = -\frac{1}{\rwo1}\intinf 
      \vci1' \philo1\, \d \myxi
      = -\frac{9\pi\sqrt{5\fal}}{50}.
\end{align*} 
Therefore, the constant $\Ctwo$ can be found from the third term, giving
\begin{align*}
 \Ctwo  = \frac{ \intinf \left(\vci1' \philo1\Ione+
\rwo1 \vci1'\philo1 -\ci \vci1' \psili1'
 -\rwi1\uci1'  \philo1\Ione \right)\, \d \xf}
{\intinf \left(\rwi1 \uci1' \philo1- \vci1' 
\philo1 \right)\, \d \xf} .
\end{align*}
Conceivably, the integrals here can be evaluated analytically, but the
results would be too complicated so in the illustrations presented
below, we have just done it numerically (remember that these are
definite integrals, so for fixed parameter values these are just
constants).  Finding $\Ctwo$ completes the derivation of $\LV1$.

For $\LV2$, the formulas \eq{pertw1fhnch6} and \eq{pertw1fhnch6p2} above
do not work as $\rw2=0$. %
Instead, we use the following expansion:
\begin{align*}
  \phil2 = \fep^{1/2} \phili2 + \dots,
  \quad
  \psil2 = \psilo2 + \fep^{1/2}\psili2 + \dots .
\end{align*}
Substituting this into the left eigenfunction equation \eq{adjlinprobfhnch6}
and balancing the powers of $\fep^{1/2}$, we have
\begin{align*}
\phili2''+\left(2\uci0-\fth\right)\phili2=0, \quad 
-\phili2+\ci\psilo2'=0, 
\end{align*}
with solutions
\begin{align}
& \phili2=-3{\fth}^{3/2}  \sech^2 \left(\xf\sqrt {\fth}/2
\right) \tanh \left(\xf\sqrt {\fth}/2 \right)/2
, \nonumber \\
& \psilo2= \frac {3\sqrt {5}{\fth}^{3/2}  \sech^2 \left(\xf 
\sqrt {\fth}/2\right)}{10\sqrt {\fal}}.
\end{align}

\sglfigure{\includegraphics{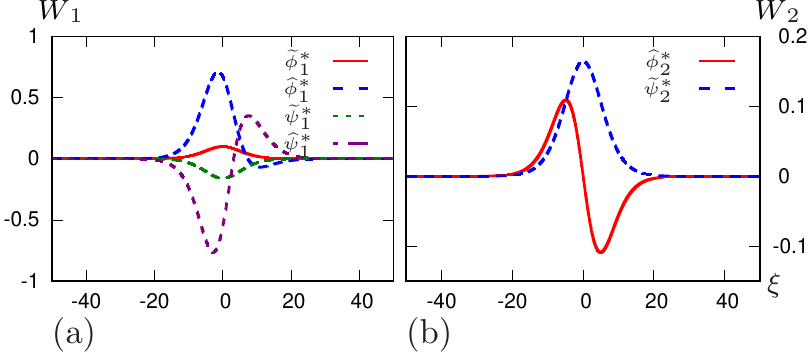}}{%
 Plot of the two components of first (a) and second (b) eigenfunctions of 
FHN system. Parameters used: $\fth=0.05$, $\fal=0.37$. 
}{fhn-perteigs}

\Fig{fhn-perteigs} shows the asymptotic components of
the first two eigenfunctions of FHN system for the parameters
$\fth=0.05$, $\fal=0.37$. The eigenfunctions vanish at $\xf\to-\infty$ by
construction; however, the result of our calculations is that they also
vanish at $\xf\to\infty$, hence it is not necessary to consider the
outer expansion for them. 

\subsection{Strength-Duration Curve}

After finding the asymptotics of the eigenfunctions of the model in
closed forms, we can use those to construct the approximation to the
critical curves. In this paper, we restrict consideration to the
strength-duration curve.
 
In our proposed procedure, the value of $\shift$ is given by the
transcendental equation \eq{zeroofs}. Employing the found asymptotics,
we obtain
\begin{align} \eqlabel{traistsch6}
\seq\left(\fth,\fep,\shift,\tst \right) \bydef & 
  2\Numtwo  \e^{\rw1 \shift/\c}\bigg[ \left(1+ \sqrt{\fep} \Ctwo\right) \Ithree\left(\fth,\shift,\tst\right)
  \nonumber  \\ & 
 -\frac { 9 \sqrt{\fep\fal}}{\sqrt{5\fth}}
\Ifour \left(\fth,\shift,\tst\right)
\bigg] \\
& + 3 \Numone \sqrt{\fep}{\fth} \Ifive \left(\fth,\shift,\tst\right)
= 0. \nonumber
\end{align}
The integral $\Ithree$ in this equation is calculated as 
\begin{align*}
\Ithree\left(\fth,\shift,\tst\right) &=
  \int\limits_{-\shift}^{-\c \tst-\shift} \e^{\rw1\xf/\c} \sech^3\left(\xf \sqrt{\fth}/2 \right)\,\d\xf \\
  &= \frac{1}{\sqrt{\fth}} \left( \Trone\left( \e^{ -\left(\c \tst+\shift\right)\sqrt{\fth}}\right)- \Trone\left( \e^{-\shift \sqrt{\fth}}\right) \right),
\end{align*}
where
\begin{align*}
\Trone\left(\rrr\right)&= {\frac {2{\rrr}^{(\a+1)/2} \left( \a\rrr+\a+\rrr-1 \right)   }{ \left( \rrr+1 \right) ^{2}}}  \\
&-{\rrr}^{(\a+1)/2} \left( \a^2-1 \right)  
 \Lerch \left( -\rrr,1,(\a+1)/2 \right), \\
\a & =\frac{2\rw1}{\c\sqrt{\fth}}, 
\end{align*}
and $\Lerch$ is the Lerch transcendent, defined \eg\ in~\refcite{bateman1955higher} as
\begin{align}
\Lerch\left(\z,\k,\q\right)= \sum_{\n=0}^{\infty} \frac{\z^\n}{\left(\q+\n\right)^\k},
\end{align}
provided that $|\z|<1$ and $\q\neq 0,-1,\ldots$.

The integral $\Ifour$ is also calculated as a function of
the Lerch transcendent as
\begin{align*}
\Ifour\left(\fth,\shift,\tst\right) &=\int\limits_{-\shift}^{-\c \tst-\shift} {\e ^{\rw1 \xf/\c} \frac {\xf{{\e}^{\xf\sqrt {\fth}}}\sech^3\left(\xf \sqrt{\fth}/2 \right)
}{  {{\e}^{\xf\sqrt {\fth}}}+1  }}\, \d\xf \\
&= \frac{2}{{\fth}} \left( \Trtwo\left( \e^{ -\left(\c \tst+\shift\right)\sqrt{\fth}}\right)- \Trtwo\left( \e^{-\shift \sqrt{\fth}}\right) \right),
\end{align*}
where 
\begin{align*}
  \Trtwo(\rrr) & = 
    \frac{{\rrr}^{(\b+1)/2}} { 6\left( \rrr+1 \right) ^{3}} \bigg[ 
      \left( 
        -\left( \b^2-2\b-3 \right) \rrr^2
  \right. \\ & \left.
        -\left( 2\b^2-6\b-8\right) \rrr - \b^2 + 4\b - 3
      \right) \ln(\rrr) 
  \\ & 
      -4\,\b{\rrr}^{2}-8\,\b\rrr+4\,{\rrr}^{2}-4\,\b+12\,\rrr+8
  \bigg]
  \\ & 
  -\frac{
    \rrr^{(\b+1)/2} (\b+1) (\b^2-4\b+3)
  } {12} \Lerch \left(-\rrr,2,(\b+1)/2 \right) 
  \\ &
  + \frac{1}{12} \left(
    \ln(\rrr) \b^3-3\b^2 \ln(\rrr) - \b\ln(\rrr) + 6\b^2
  \right. \\ &  \left.
    + 3\,\ln(\rrr) - 12\b - 2 
   \right) \rrr^{(\b+1)/2} \Lerch\left( -\rrr,1,(\b+1)/2\right), \\
  \b &=2\left(\frac{\rw1}{\c\sqrt{\fth}}+1\right) 
\end{align*}
and finally the  integral $\Ifive$ is calculated as
\begin{align*}
\Ifive\left(\fth,\shift,\tst\right) & = \int\limits_{-\shift}^{-\c \tst-\shift}  \sech^2 \left(\xf\sqrt {\fth}/2
\right) \tanh \left( \xf\sqrt {\fth}/2 \right)\,
 \d \xf \\
 & = {\frac {  \sech^2 \left(\shift\sqrt {\fth}/2\right) - \sech^2 \left(\left( \c{\tst}+ \shift \right) \sqrt {
\fth}/2\right) }{\sqrt {\fth}}}.
\end{align*}

A further simplification can be achieved by taking into account that
$\fth$ is also a small parameter. With a substitution
$\rrr=\e^{\myxi\sqrt{\fth}}$, the limits of all three integrals
become close to $1$. Hence, these integrals can be evaluated as the
Taylor expansion around $1$ and they become regular functions,
\begin{align*}
 \Ithree\left(\fth,\shift,\tst\right)&= \frac{8}{\sqrt{\fth}} 
\int\limits_{\e^{-\shift \sqrt{\fth}}}^{\e^{ -\left(\c \tst+\shift\right)\sqrt{\fth}}}  \frac{\rrr ^{\left(\a+1\right)/2}}{\left(\rrr+1\right)^3}\,\d \rrr \\
& \approx \frac{1}{\sqrt{\fth}} \int\limits_{ 1- \shift \sqrt{\fth}}^{1 -\left(\c \tst+\shift\right)\sqrt{\fth}} \left(1+ \left( \rrr-1 \right)  \left( \a/2-1 \right) \right)\,\d \rrr \\
&= \rw1 \tst\shift+ \fep^{1/2} \ci\tst\ \left( \rw1\tst/2-\shift\sqrt {\fth}-1 \right)\\
&-\fep\ci^2\tst^2\sqrt {\fth}/2,
\end{align*}
\begin{align*}
 \Ifour\left(\fth,\shift,\tst\right)&= \frac{16}{{\fth}} 
\int\limits_{\e^{-\shift \sqrt{\fth}}}^{\e^{ -\left(\c \tst+\shift\right)\sqrt{\fth}}}  \frac{ \ln\left(\sqrt{\rrr}\right) \rrr ^{ \left(b+1\right)/2}}{\left(\rrr+1\right)^4}\,\d \rrr \\
& \approx \frac{1}{\fth} \int\limits_{ 1- \shift \sqrt{\fth}}^{1 -\left(\c \tst+\shift\right)\sqrt{\fth}} \left(\rrr-1\right)\,\d \rrr \\
&=\fep^{1/2} \ci\tst\shift+\fep \ci^2\tst^2/2,
\end{align*}
\begin{align*}
 \Ifive\left(\fth,\shift,\tst\right)&= \frac{4}{\sqrt{\fth}} 
\int\limits_{\e^{-\shift \sqrt{\fth}}}^{\e^{ -\left(\c \tst+\shift\right)\sqrt{\fth}}}  \frac{\rrr-1}{\left(\rrr+1\right)^3}\,\d \rrr \\
&\approx \frac{1}{2\sqrt{\fth}} \int\limits_{ 1- \shift \sqrt{\fth}}^{1 -\left(\c \tst+\shift\right)\sqrt{\fth}} \left(\rrr -1 \right)\,\d \rrr \\
& =\fep^{1/2} \ci\tst\shift\sqrt {\fth}/2+\fep \ci^2\tst^2\sqrt {\fth}/4.
\end{align*}
Plugging these back into the transcendental equation \eq{traistsch6}, we have
\begin{align*}
& \Numtwo  \e ^{\rw1 \shift/\c} \left(1+ \fep^{1/2}\Ctwo\right) \left(\rw1 \tst\shift+ \fep^{1/2} \ci\tst\
\right. \\
&  \left. \left(\rw1\tst/2-\shift\sqrt {\fth}-1 \right)-\fep\ci^2\tst^2\sqrt {\fth}/2\right)\\
& - 
 \frac { 9 \sqrt{\fal} \Numtwo  \e ^{\rw1 \shift/\c}}{\sqrt{5\fth}}\left(\fep \ci\tst\shift+\fep^{3/2}\ci^2\tst^2/2\right) \\
 & +\frac{3\Numone{\fth}^{3/2}}{4} \left(\fep \ci\tst\shift\sqrt {\fth}+\fep^{3/2} \ci^2\tst^2\sqrt {\fth}/2\right) =0, 
\end{align*}
and equating this up to the order of $\fep^{1/2}$ gives the value of $\shift$ as
\begin{align}
\shift=\frac{\ci\left(2-\rw1\tst\right)}{2\left(\rw1 \Ctwo-\ci\sqrt {\fth}\right)}+\mathcal{O}\left(\fep\right).
\end{align}
Having obtained a closed expression for $\shift$, the final step is to
substitute this into one of the equations in \eq{crit-sys}, which
finally delivers the formula for the strength-duration curve:
\begin{align}\eqlabel{asymistsch6}
\ampi &=\frac{\Numone}{2\int\limits_0^{\tst}\e ^{-\rw1 \tfs}\phil1\left(-\c\tfs-\shift\right)\,  \d {\tfs}} \\
&= \frac{ -\c \Numone}{2 \e ^{\rw1 \shift/\c}\left[ \left(1+ \sqrt{\fep}\Ctwo\right) \Ithree\left(\fth,\shift,\tst\right) -\frac { 9 \sqrt{\fep\fal}}{\sqrt{5\fth}}
\Ifour \left(\fth,\shift,\tst\right)
\right]}.\nonumber
\end{align}

\sglfigure{\includegraphics{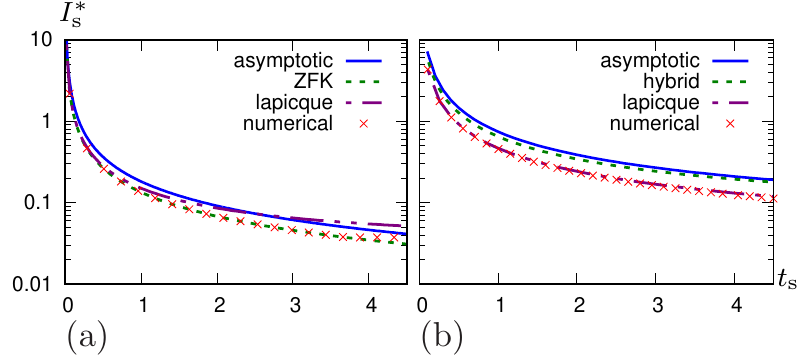}}{%
  (a) Sketch of the comparison between analytical and numerical
  strength-duration curve where we used the perturbation analysis for
  analytical derivation and Lapicque and ZFK curves also plotted for
  the parameters $\fth=0.01$, $\fal=0.37$, $\fep=10^{-5}$. %
  (b) Same for $\fth=0.05$, $\fal=0.37$, $\fep=10^{-2}$, apart from we
  do not include ZFK result and add hybrid approach analytical
  derivation instead. %
}{fhn-pertistsch6}

The plots of the asymptotic threshold curves given by
  equation \eq{asymistsch6}, compared against the direct numerical
  simulations, are shown in \Fig{fhn-pertistsch6}. The left panel of
  the figure shows the case of a very small $\fep$, and we show also the
  numerical curve for ZFK equation, \ie\ the $\fep\to0$ limit of FHN.
  We observe that there is a good agreement between the two numerical
  curves.  In the right panel of the figure, the values of both $\fep$
  and $\fth$ are increased compared to the left panel.  Here instead
  of the ZFK curve, we show the ``hybrid'' numeric-asymptotic
  prediction, that is, the asymptotic result given by linearized
  theory \eqthree{crit-sys,Numnboth,zeroofs}, in which the ingredients
  $\buc$, $\LV{1,2}$ and $\rw1$ are found numerically using methods
  described in~\refcite{bezekci2015semianalytical}.  The closeness
  of the asymptotic curve to the hybrid curve in this panel,
  despite the fact that $\fep$ and $\fth$ are not very small, is an
  illustration of the quality of the asymptotic formulas for the said
  ingredients, which is the main technical result of this paper.
  Expectedly, the asymptotic threshold curve, in this case, is not
  better than the hybrid prediction.

Out of curiosity, for each figure we also plot the
  curves described by the equation 
\begin{align}\eqlabel{LBH}
  \ampi=\frac{\IRH}{1-\exp\left(-\tst/ \CHRO\right)} ,
\end{align}
which is the classical formula going back to works
by~\textcite{lapicque1907recherches}, \textcite{blair1932intensity}
and \textcite{hill1936excitation} as a phenomenological law
approximating excitation thresholds in a wide range of
electrophysiological experiments, well before the realistic models of
any biological excitable tissues became available. For theoretical
justification, already~\textcite{lapicque1907recherches} proposed a
hypothetical linear electric circuit that can produce this dependence;
nowadays this may be considered as a linearization of actual nonlinear
membrane equations. In the spatially extended context, it has been
shown\cite{idris2008analytical} that~\eq{LBH} automatically emerges as
a result of the linearized theory \eqthree{crit-sys,Numnboth,zeroofs}
in the case of ``critical nucleus'', $\c=0$, $\shift=0$, regardless of
other details of the model.
In~\fig{fhn-pertistsch6}, the values of the rheobase $\IRH$ and
chronaxie $\CHRO$ are not obtained theoretically, but fitted to the
numerical curves using the Levenberg-Marquardt nonlinear
least-squares fitting
algorithm~\cite{levenberg1944method,marquardt1963algorithm}.  We note
that the Lapicque-Blair-Hill curve fits the results of direct
simulations much better for the right panel, even though the
theory\cite{idris2008analytical} promises  its applicability to the
case $\fep=0$, which is closer to the case in the left panel. The
reasons for this paradoxical discrepancy, apart from the simple fact
that the analytical formula is only approximate in any case, are not
clear at present and require further investigation.

\section{Discussion}
\seclabel{disc}

The semi-analytical approach to the strength-extent and
strength-duration threshold curves has been presented in our previous
publications\cite{bezekci2015semianalytical,bbvnbists}. In the
multicomponent reaction-diffusion systems, the essential ingredients for
the case of the strength-extent curve are moving critical solution
(either critical front or critical pulse) and two leading left (adjoint)
eigenfunctions, whereas for the case of the strength-duration curve, we
additionally need the positive eigenvalue $\rw1$.  In
\refcites{bezekci2015semianalytical} and [\onlinecite{bbvnbists}], the case of
FitzHugh-Nagumo was considered among others, and these ingredients were
found only numerically, which of course depreciated the heuristical value
of the results, not to speak of associated computational cost and
numerical analyst's effort.

The main aim of this article has been to overcome this disadvantage to
approximately calculate the analytical expressions for the ingredients
of the FHN theory and obtain a closed-form expression for the critical
curve. As FHN system is considered as a ZFK equation extended by a slow
variable and all essential ingredients of ZFK equation are known explicitly
in the limit of small $\fth$, it is possible, therefore, that the
perturbation theory can be applied in a straightforward way to determine
all essential ingredients of the FHN system, and even hence the critical
curve itself analytically. 

An example of a qualitative result afforded by the fully analytical
approach, is the deviation of the strength-duration curve from the
classical Lapique-Blair-Hill formula%
\cite{lapicque1907recherches,blair1932intensity,hill1936excitation}. %
It has been noted that in some cardiac excitation models this formula
requires adjustments in order to fit the experimental or numerical
curves, see \eg~\refcite{Noble-Stein-1966}, where this deviation has
been associated with the phenomenon of the membrane accommodation
(described by the slow variable in FHN). In the context of the ignition
problem in a spatially extended system, the accommodation is manifested
by the fact that the critical solution is not a stationary ``critical
nucleus'', but a propagating
solution\cite{idris2007critical,bezekci2015semianalytical}, and it is a
rather general result of~\refcite{idris2008analytical} that critical
nucleus implies Lapique-Blair-Hill strength-duration dependency, at
least in the linear approximation. Hence an example with accommodation
where moving critical solution and corresponding strength-duration curve
can be described in a closed form, is an important step in understanding
of how accommodation affects the threshold properties of spatially
extended excitable systems.

Some obvious extension of our approach is to generalize for different
temporal profiles of the stimulating current, and also for other
initiation protocols, such as stimulation by voltage (strength-extent
curve). It also would be interesting to investigate the feasibility of
using perturbation theory on some more realistic cardiac excitation
models with larger number of dynamical variables, in which case the
computational cost of the essential ingredients increases.

\begin{acknowledgments}
VNB gratefully acknowledges the current
financial support of the EPSRC via grant EP/N014391/1 (UK)
\end{acknowledgments}

%

\end{document}